# Probing restricted diffusion and exchange using free gradient waveforms: validation by numerical simulations

Arthur Chakwizira[1], Carl-Fredrik Westin[2], Jan Brabec[3], Samo Lasič[4,5], Linda Knutsson[1,6,7], Filip Szczepankiewicz[3] and Markus Nilsson[3]

1. Medical Radiation Physics, Lund, Lund University, Lund, Sweden
2. Department of Radiology, Brigham and Women's hospital, Harvard Medical School, Boston, MA, United States
3. Clinical Sciences Lund, Lund University, Lund, Sweden
4. Danish Research Centre for Magnetic Resonance, Centre for Functional and Diagnostic Imaging and Research, Copenhagen University Hospital - Amager and Hvidovre, Copenhagen, Denmark
5. Random Walk Imaging AB, Lund, Sweden
6. Russell H. Morgan Department of Radiology and Radiological Science, Johns Hopkins University School of Medicine, Baltimore, MD, United States
7. F. M. Kirby Research Center for Functional Brain Imaging, Kennedy Krieger Institute, Baltimore, Maryland, USA.

**Corresponding author:**
Arthur Chakwizira, Department of Medical Radiation Physics, Lund University, Skåne University Hospital, SE-22185 Lund, Sweden
Email address: arthur.chakwizira@med.lu.se

**Submitted to:** NMR in Biomedicine

**Word count:** approximately 6700

**Sponsors/Grant numbers:**

- VR (Swedish Research Council)
  - 2016-03443
  - 2020-04549
- eSSENCE
  - 6:4
- Cancerfonden (Swedish Cancer Foundation)
  - 2019/474

**Keywords:** diffusion MRI, time-dependence, restricted diffusion, exchange, gradient waveform, restriction-exchange space, cumulant expansion, restriction-exchange weighting

**Abbreviations:** dMRI (diffusion MRI), FWF (Free waveforms), SDE (Single Diffusion Encoding), DDE (Double Diffusion Encoding), OGSE (Oscillating Gradient Spin Echo), PGSE (Pulsed Gradient Spin Echo), FEXI (Filter Exchange Imaging), AXR (Apparent Exchange Rate), CTI (Correlation Tensor Imaging), IMPULSED (Imaging Microstructural Parameters Using Limited Spectrally Edited Diffusion), POMACE (Pulsed and Oscillating gradient MRI for Assessment of Cell size and Extracellular space), VERDICT (Vascular, Extracellular, and Restricted Diffusion for Cytometry in Tumours)

# Abstract

Monitoring time-dependence with diffusion MRI yields observables sensitive to compartment sizes (restricted diffusion) and membrane permeability (water exchange). However, restricted diffusion and exchange have opposite effects on the diffusion-weighted signal, which can confound parameter estimates. In this work, we present a signal representation that captures the effects of both restricted diffusion and exchange up to second order in b-value and is compatible with gradient waveforms of arbitrary shape. The representation features mappings from a gradient waveform to two scalars that separately control the sensitivity to restriction and exchange. We demonstrate that these scalars span a two-dimensional space that can be used to choose waveforms that selectively probe restricted diffusion or exchange, in order to eliminate the correlation between the two phenomena. We found that waveforms with specific but unconventional shapes provide an advantage over conventional pulsed and oscillating gradient acquisitions. We also show that parametrisation of waveforms into a two-dimensional space can be used to understand protocols from other approaches that probe restricted diffusion and exchange. For example, we find that the variation of mixing time in filter-exchange imaging corresponds to variation of our exchange-weighting scalar at a fixed value of the restriction-weighting scalar. Numerical evaluation of the proposed signal representation using Monte Carlo simulations on a synthetic substrate showed that the theory is applicable to sizes in the range 2 – 7 μm and barrier-limited exchange in the range 0 – 20 $s^{-1}$. The presented theory constitutes a simple and intuitive description of how restricted diffusion and exchange influence the signal as well as how to design a protocol to separate the two effects.

# 1 Introduction

Diffusion MRI (dMRI) is an important radiological tool because it sensitizes the MR signal to the diffusion of water molecules, which indirectly probes tissue microstructure [1–6]. Today, dMRI provides clinically useful information about the microstructure via the apparent diffusion coefficient (ADC) [7–9]. Although the ADC is sensitive to pathology, it is not specific. This can be remedied by probing additional information pertaining to for example the time-dependence of the diffusion process [10,11]. Particularly, sensitizing the dMRI experiment to restricted diffusion and exchange enables inference of metrics sensitive to cell sizes and membrane permeability, respectively. However, restricted diffusion and exchange have opposite effects on the diffusion-weighted signal [12,13]; as the diffusion time increases, the signal for a given encoding strength (b-value) is elevated by restricted diffusion and reduced by exchange. Therefore, the estimation of, for example, cell sizes, can be confounded by a high membrane permeability, and vice versa. Current approaches often assume that restricted diffusion and exchange dominate in different time regimes; the former at short time scales and the latter at long time scales [11]. This assumption, if not violated, enables straightforward estimation of one of the parameters while neglecting the other. Some approaches have on the other hand incorporated both restricted diffusion and exchange [11,14–16]. These were based on the use of the standard pulsed gradient spin echo (PGSE) or the oscillating gradient spin echo (OGSE) sequences for the diffusion encoding. However, our preliminary work[13] indicated that more efficient measurements of restricted diffusion and exchange can be achieved with gradient waveforms that can be arbitrarily modulated. This work explores when and how such benefits manifest.

Cell sizes can be estimated with dMRI in numerous ways. Early approaches used the diffraction-like signal patterns in the narrow pulse limit to estimate restriction lengths in materials such as polystyrene spheres [17] and erythrocytes [18]. However, *in vivo* applications are challenging since exceptionally strong gradients are required to probe small structures [19,20], and these systems are rarely used in human imaging [21]. Distributions of cell sizes can also obscure diffraction patterns [22–25]. More recent approaches address this by instead utilizing compartment models that describe cells as impermeable objects in an extracellular medium [11,26,27]. Examples applied for cancer imaging include IMPULSED [28], POMACE [29,30] and VERDICT [31]. IMPULSED quantifies compartment sizes and the intracellular fraction using both OGSE and PGSE acquisitions. POMACE also exploits both PGSE and OGSE measurements in different timescales to estimate the extracellular volume fraction, cell size and compartment diffusivities. VERDICT also uses PGSE, but incorporates a vascular compartment modelled by an anisotropic diffusion tensor. Sizes can also be estimated without using explicit tissue models. Nilsson et al. [32] proposed an approach in which a Taylor series approximation of the diffusion spectrum is used to capture restriction-driven time-dependent diffusion. The approach was sensitive to restrictions larger than 4 μm using gradient strengths available at clinical MRI systems. Notably, the approach is also distinguished by its ability to accommodate diffusion encoding by arbitrary gradient waveforms.

Membrane permeability (quantified by the water exchange rate) has also been estimated in various ways. Latour et al. [33] proposed an effective medium theory describing the effect of permeability on diffusion in the long-time regime. The approach was applied to estimate membrane permeability in bovine red blood cells [33] and yeast [34]. Novikov et al. [14] treated cell membranes as randomly oriented flat layers to define a forward model that incorporated effects of permeability. This model was recently applied to estimate cell permeability and size *in vivo* [35]. All these approaches employ a pair of pulsed gradients for the diffusion encoding. However, using two pairs of pulsed gradients with a variable mixing time—referred to as double diffusion encoding [36]—can be beneficial for exchange mapping [37,38]. Lasič et al. [39] proposed filter exchange imaging (FEXI), which is sensitive to cell permeability captured by the so-called apparent exchange rate (AXR). This parameter can differentiate between viable and necrotic parts of a tumour [40] and has been applied to characterize breast tumours [41]. Exchange-related approaches were based on pulsed gradients until Ning et al. [42] derived an approach that accommodates arbitrary gradient waveforms. The potential utility of such waveforms for exchange mapping has, however, not yet been systematically explored.

Sensitivity and specificity to tissue microstructure are governed by the experimental approach, here defined as the set of gradient waveforms used to acquire the dMRI data. As indicated in the previous sections, several waveform designs have been proposed and applied in combination with mapping of restricted diffusion and exchange. The most commonly used waveform—here referred to as single diffusion encoding (SDE) [36]—comprises a pair of identical pulsed gradients inserted in a spin-echo experiment [43]. Although this approach is the main workhorse of dMRI, more complex waveforms can have benefits. For example, DDE which uses two pairs of pulsed gradients [44] can be used to better map exchange [17,45] or restricted diffusion [46] whereas oscillating gradient waveforms have been suggested for probing diffusion at shorter time scales [47–50]. Both DDE and OGSE are fully specified by a few parameters such as pulse duration, pulse separation and oscillation frequency. While this enables closed-form expressions for the signal, it forgoes more intricate waveform designs which can increase efficiency or the amount of information encoded into the signal [51]. Diffusion encoding by free gradient waveforms (FWF) was first explored by Callaghan [52] who adopted the multiple propagator method [53] and proposed a forward model for encoding restricted diffusion with arbitrary waveforms. More recently, Drobnjak et al. [54] optimized waveforms for maximized sensitivity to restricted diffusion. Waveforms have also been optimized for efficient tensor-valued encoding and suppression of various artefacts and motion encoding [51]. Although waveform designs exist to emphasise the effects of either restricted diffusion or exchange separately, a unified framework capturing both phenomena has not yet been proposed for gradient waveforms with arbitrary shape.

In this work, we propose a signal representation describing both restricted diffusion and water exchange. We arrive at this representation using a second-order cumulant expansion of the signal attenuation due to diffusion-induced phase dispersion. We highlight that the effect of the waveform on the second and fourth cumulants of the phase distribution can be evaluated in terms of a scalar quantity that indicates how much a given gradient waveform encodes for restricted diffusion [32]. Similarly, we note that the fourth cumulant features a scalar quantity informative of how strongly a gradient waveform encodes water exchange [42]. The resulting representation is a unification of the restriction and exchange frameworks from previous studies [32,42]. We employ the waveform-dependent scalar parameters to characterise the restriction- and exchange-weighting properties of different types of waveforms. Guided by this information, we select the protocol that maximises the statistical leverage in estimates of restriction length and exchange rate. We also investigate the value of free waveforms compared to designs based on SDE, DDE and OGSE. Using Monte Carlo simulations on a regular two-compartment 2D substrate of circular cells, we also identify the interval of validity of the proposed representation. Finally, we apply the approach to analyse previously proposed experimental approaches to evaluate the degree to which they encode for restricted diffusion and exchange.

# 2 Theory

In the following, we outline the proposed restriction-exchange theory. For simplicity, only the one-dimensional case is considered (that is, diffusion and encoding in one direction at a time).

## 2.1 The restriction-exchange signal representation

We begin by introducing the signal representation that captures the effects of restricted diffusion and exchange on the normalized diffusion-weighted signal ($S/S_0$):

$$\ln(S/S_0) \approx -b \cdot [E_D + V_\omega E_R] + \frac{1}{2} b^2 \cdot [V_D + V_\omega C_{DR} + V_\omega^2 V_R] \cdot (1 - k\Gamma). \quad (1)$$

The three parameters $b$, $V_\omega$, and $\Gamma$ describe the effect of the gradient waveform and quantify the strength of the diffusion encoding, restriction encoding, and exchange encoding, respectively. The other six parameters relate to the microstructure and are described in the following. In the first-order term, we have two parameters: $E_D = \langle D \rangle$, which is the mean diffusivity where $D$ is the compartment-specific long-time apparent diffusivity, and $E_R = \langle R \rangle$ which is the average restriction coefficient that in environments with restricted diffusion is given by $R = cd^4/D_0$ where $d$ is an index sensitive to the restriction length (compartment diameter), $D_0$ is the bulk diffusivity and $c$ is a geometry-dependent constant. The operator $\langle \cdot \rangle$ denotes averaging across environments. Furthermore, the second order term features another four parameters: $V_D$, which is the variance in $D$, $C_{DR}$ is the covariance between $D$ and $R$, $V_R$ is the variance of $R$, and finally $k$ which is the average water exchange rate between different environments. A derivation of the representation and a motivation of these terms follows below.

## 2.2 Deriving the restriction-exchange signal representation

Equation 1 can be deduced by expressing the signal attenuation due to diffusion in terms of the spin phase distribution and its cumulant expansion up to fourth order,

$$\frac{S}{S_0} = \langle \exp(-i\phi) \rangle \approx \exp\left(-\frac{1}{2}\langle \phi^2 \rangle + \frac{1}{24}(\langle \phi^4 \rangle - 3\langle \phi^2 \rangle^2)\right), \quad (2)$$

where $\langle \cdot \rangle$ is an average over all spin trajectories contributing to the signal. The phase is defined as the inner product of the gradient waveform, $g(t)$, and the spin trajectory, $r(t)$,

$$\phi = \gamma \int_0^T g(t) \cdot r(t) \mathrm{d}t, \quad (3)$$

where $\gamma$ is the gyromagnetic ratio. Together, Eq. 2 and 3 provide a linear description of the signal as a sum of inner products between two continuous functions: one is the encoding function $g(t)$ and the other is the function carrying microstructure information $r(t)$. We now seek functions that map the continuous encoding function $g(t)$ onto a set of scalar weights that control the sensitivity of the cumulants to restricted diffusion and to water exchange, that is, $V_\omega$, and $\Gamma$. We proceed by evaluating the cumulants assuming a scenario where spins can exist in distinct environments in which some exhibit approximately Gaussian diffusion and others restricted diffusion.

## 2.3 Evaluating the second cumulant

The second cumulant of the phase distribution is given by [55]

$$\frac{1}{2}\langle \phi^2 \rangle = \frac{1}{2\pi}\int_{-\infty}^{\infty} |q(\omega)|^2 D(\omega)\, \mathrm{d}\omega, \quad (4)$$

where $D(\omega)$ is the spectrum of diffusion coefficients and $|q(\omega)|^2$ is the diffusion encoding power spectrum given by

$$|q(\omega)|^2 = \left|FT\big(q(t)\big)\right|^2, \quad (5)$$

where $FT$ denotes Fourier transform and $q(t)$ is the dephasing q-vector related to the gradient waveform via

$$q(t) = \gamma \int_0^t g(\tau)\mathrm{d}\tau. \quad (6)$$

At low frequencies, the diffusion spectrum can be represented by its Taylor series expansion [32,55–57] which we express as

$$D(\omega) \approx \beta_0 + \beta_2 \omega^2, \tag{7}$$

where $\beta_n = D^{(n)}(0)/n!$.

Equation 4 can describe the signal from the whole voxel or the signal from one microenvironment. For environments in which the diffusion is Gaussian ($D(\omega) = D$) we have $\beta_0 = D$ and $\beta_2 = 0$, where $D$ is the (apparent) diffusion coefficient for that environment. For environments in which the diffusion is restricted, we have $\beta_0 = 0$ and $\beta_2 = R$ [32,57]. Given diffusion inside certain geometries (cylinders and spheres), the so-called restriction coefficient $R$ has the general form [32,55,57]

$$R = cd^4 D_0^{-1}, \tag{8}$$

where $d$ represents the compartment diameter, $D_0$ is the bulk diffusivity and $c$ is a constant with the value 7/1536 for a cylinder and 1/500 for a sphere. We will refer to the parameter $d$ as merely "size", indicating that it is sensitive to, but not necessarily equal to, the compartment diameter.

The second cumulant for the whole voxel can now be written

$$\frac{1}{2}\langle\phi^2\rangle = b\,(\langle\beta_0\rangle + V_\omega \langle\beta_2\rangle). \tag{9}$$

where $\langle\beta_0\rangle$ and $\langle\beta_2\rangle$ are averages across microenvironments. The two experiment-related parameters are given by

$$b = \frac{1}{2\pi}\int_{-\infty}^{\infty} |q(\omega)|^2 \,\mathrm{d}\omega = \int_0^T q^2(t)\mathrm{d}t \tag{10}$$

and [32]

$$V_\omega = \frac{1}{2\pi b}\int_{-\infty}^{\infty} |q(\omega)|^2 \omega^2 \,\mathrm{d}\omega = \frac{\gamma^2}{b}\int_0^T g^2(t)\mathrm{d}t. \tag{11}$$

The latter equation provides a mapping from the gradient waveform to a scalar weight that controls time-dependent restriction encoding ($V_\omega$), which can be interpreted as the variance of the encoding power spectrum. Note that $V_\omega$ is independent of the b-value.

In a voxel with mixed environments where diffusion is approximately Gaussian in some and restricted in others, we note that the second cumulant is independent of exchange [42,58], at least as long as the exchange is barrier limited [40,59]. This means that the voxel-averaged second cumulant is simply given by

$$\frac{1}{2}\langle\phi^2\rangle = -b \cdot \mathrm{ADC}, \tag{12}$$

where the apparent diffusion coefficient is given by the environment-averaged values of $\beta_0$ and $\beta_2$ such that

$$\mathrm{ADC} = \langle\beta_0\rangle + V_\omega \langle\beta_2\rangle = E_D + V_\omega E_R \tag{13}$$

where $E_D = \langle\beta_0\rangle$ and $E_R = \langle\beta_2\rangle$. Note that in a two-compartment system with equal amounts of signal coming from environments with Gaussian diffusion and restricted diffusion, we would find $E_D = D/2$ and $E_R = R/2$, where $D$ is the diffusion coefficient of the Gaussian environment and $R$ is the restriction coefficient of the environment with restricted diffusion.

## 2.4 Evaluating the fourth cumulant

In the absence of exchange, the fourth cumulant is given by

$$\frac{1}{12}(\langle\phi^4\rangle - 3\langle\phi^2\rangle^2) = \frac{1}{2}b^2\mathrm{V(ADC)}, \quad (14)$$

where the variance in apparent diffusivities across environments is defined as [60]

$$\mathrm{V(ADC)} = \mathrm{V}(\beta_0 + V_\omega\beta_2) = V_D + 2V_\omega C_{D,R} + V_\omega^2 V_R, \quad (15)$$

where $V_D = \langle\beta_0^2\rangle - \langle\beta_0\rangle^2$ and $V_R = \langle\beta_2^2\rangle - \langle\beta_2\rangle^2$ represent the variance in free-diffusion and restriction coefficients, respectively, while $C_{D,R} = \langle\beta_0\beta_2\rangle - \langle\beta_0\rangle\langle\beta_2\rangle$ denotes the covariance between the two. Note that equation 14 assumes zero intra-environmental kurtosis [61].

Exchange reduces the variance in apparent diffusivities [42], which means the fourth cumulant is modified according to

$$\frac{1}{12}(\langle\phi^4\rangle - 3\langle\phi^2\rangle^2) = \frac{1}{2}b^2\mathrm{V(ADC)}\cdot h(k), \quad (16)$$

where

$$h(k) = \int_0^T \widetilde{q_4}(t)\exp(-kt)\,\mathrm{d}t. \quad (17)$$

Here $k$ is the exchange rate and $\widetilde{q_4}(t) = q_4(t)/b^2$ where $q_4(t)$ is the fourth-order autocorrelation function of the dephasing q-vector given by

$$q_4(t) = \int_0^T q^2(t')q^2(t'+t)\mathrm{d}t'. \quad (18)$$

We refer to $q_4(t)$ as the exchange-weighting function. Under the approximation $\exp(-kt) \approx 1 - kt$, the function $h(k)$ reduces to

$$h(k) = 1 - k\Gamma, \quad (19)$$

where

$$\Gamma = \int_0^T \widetilde{q_4}(t)\,\mathrm{d}t \quad (20)$$

and is independent of the b-value. $\Gamma$ is our second mapping from the gradient waveform to a scalar weight, which in this case controls exchange encoding via the coupling $k\Gamma$.

## 2.5 Assumptions and their implications

There are six primary assumptions to consider. First, the restriction-related parameter, $V_\omega$, was defined by approximating the diffusion spectrum with a second order Taylor series expansion (Eq. 7). Consequently, approximation errors will appear with increasing frequency of the gradient waveform or larger sizes. The error can be characterised by defining a metric that captures the relative amount of encoding power at high frequencies:

$$\eta_d = \frac{\int_{\omega_{thr}(d)}^{\infty} |q(\omega)|^2 \mathrm{d}\omega}{\int_0^{\infty} |q(\omega)|^2 \mathrm{d}\omega} \ , \qquad (21)$$

where $\omega_{\text{thr}}(d)$ is a threshold frequency above which the $\omega^2$ approximation ceases to correctly represent the diffusion spectrum for a compartment of size $d$. The threshold frequency in the calculation of $\eta_d$ was defined as the frequency at which the relative difference between the full diffusion spectrum and the Taylor series expansion exceeded 10%. An illustration of this can be found in Fig. A2 of Appendix A. Second, the derivation of $V_\omega$ assumes either spheres, cylinders or parallel planes [55] but not irregular geometries such as undulating fibres which do not necessarily feature quadratic low-frequency spectra [62]. Third, the exchange-weighting parameter, Γ, was defined using a first-order approximation (Eq. 19), which means it is valid only for relatively slow exchange or weak exchange-weighting (small Γ). The extent to which a given gradient waveform respects this first-order approximation at a given exchange rate can be quantified by defining the metric $\eta_k$ that compares the full exchange-weighting term $h(k)$ in Eq. 17 and its approximation in Eq. 19:

$$\eta_k = (h(k) - (1 - k\Gamma))^2. \qquad (22)$$

Since Γ is the first moment of the exchange-weighting function ($q_4$), waveforms whose $q_4$ rapidly decays to zero will have a smaller $\eta_k$. Fig. A1 of Appendix A depicts this comparison for three waveforms: one with short total encoding time ($T$) and small Γ and two with the same encoding time but different Γ. Fourth, the theory assumes barrier-limited exchange—a requirement which can bias estimates if not fulfilled. To illustrate, if $\tau$ is the residence time and $t_c$ the position correlation time, exchange is barrier-limited when $t_c \ll \tau$ [40,59]. Therefore, some combinations of large sizes and short exchange times are expected to violate this fourth assumption. Fifth, the exchange weighting term in Eq. 17 was derived for two Gaussian components [42] and generalisation to non-Gaussian diffusion is here merely assumed. However, the quality of this assumption will be analysed. Sixth and finally, Eq. 1 assumed low to moderate b-values because the cumulant expansion is performed around a b-value of zero up to the second order.

# 3 Methods

## 3.1 Characterization of waveforms in the restriction-exchange space

The theory highlighted two parameters, $V_\omega$ and Γ, that describe how strongly a gradient waveform encodes for restricted diffusion and exchange, respectively. The first goal was to characterize these parameters for a large set of waveforms. Four waveform families were investigated: pulsed waveforms (SDE and DDE), oscillating waveforms (OGSE), and free waveforms (FWF). Waveforms were generated for each family and then mapped to the restriction-exchange space (plot of $V_\omega$ vs Γ), which was then analysed.

SDE waveforms were generated by varying the pulse width (δ) and the pulse separation (Δ) as described in Appendix B. DDE waveforms were generated by varying the mixing time, the leading-edge separation ($\Delta_1, \Delta_2$) and the pulse width (δ), while assuming symmetric DDE where $\delta_1 = \delta_2$ and $\Delta_1 = \Delta_2$. Pulse shapes were trapezoidal with the ramp time determined by the amplitude of the waveform and the maximum slew rate. OGSE waveforms included both sine- and cosine-modulated varieties, and we varied the frequency of oscillation, the pulse duration (δ) and the pause duration for an RF pulse (Δ − δ). For FWF, a large set of waveforms aimed at spanning the set of executable waveforms was generated by the procedure described in Appendix B. All waveforms were subjected to the following constraints:

1. The spin-echo condition where $\int_0^T g(t)\,\mathrm{d}t = 0$
2. Maximum gradient amplitude of 80 mT/m at a b-value of 5 ms/μm$^2$
3. Maximum slew rate of 70 T/m/s
4. Minimum pause duration of 9 ms [63]
5. Maximum total encoding time of 250 ms (shorter times were allowed)
6. Unique combination of $\Gamma$ and $V_\omega$ (that is, values distinct from those of all other generated waveforms of the same type)
7. Symmetric about the 180° refocusing pulse to automatically satisfy the spin-echo condition and avoid concomitant gradient effects [64].

Note that there can be multiple waveforms mapping to approximate the same point in the restriction-exchange space. These may have different efficiencies, defined as the b-value attained over a given total encoding time and maximal gradient amplitude [65,51]. We accounted for this by selecting the most efficient candidate out of 100 generated waveforms in every neighbourhood of the space.

The full set of waveforms for all four families (SDE, DDE, OGSE, FWF) were analysed in terms of the region spanned in the restriction-exchange space, as well as in terms of how well they complied with the assumptions highlighted in section 2.2. For this analysis we employed the two metrics $\eta_d$ and $\eta_k$ defined in Eq. 21 and 22. For each waveform family, we identified the waveforms satisfying $\eta_d \leq 70\%, \eta_d \leq 30\%$ and $\eta_d \leq 10\%$ as well as $\eta_k \leq 70\%, \eta_k \leq 30\%$ and $\eta_k \leq 10\%$. The metric $\eta_d$ was computed at a diameter of 20 μm and $\eta_k$ at an exchange rate of 20 s$^{-1}$. These values were chosen because they comprised the upper bound in the simulations (described later).

## 3.2 Investigating different protocols for fixed microstructure parameters

Having characterized waveforms in the restriction-exchange space, we proceeded to evaluate the performance of the representation in Eq. 1 for a fixed microstructure but varying protocols. For this purpose, a set of protocols each comprising four waveforms was defined (details given below). Synthetic measurements were then performed using Monte Carlo simulations (described in Appendix C). Model parameters were estimated from the simulated signals by fitting Eq. 1 using the non-linear least squares solver *lsqnonlin* in MATLAB (The MathWorks, Natick, MA, R2019a). Four parameters were fitted ($E_D, E_R, V_D$ and $k$) while $V_R$ and $C_{D,R}$ were fixed at zero. The rationale for fixing the last two parameters was that the variance term in the cumulant expansion was dominated by $V_D$ for all waveforms and simulation substrates considered in this work (data not shown).

The size index was computed from the fitted $E_{\mathrm{R}}$ using $d = \left(D_{\mathrm{in}} \cdot E_{\mathrm{R}}/(c \cdot f_{\mathrm{in}})\right)^{1/4}$and the known underlying values of the intracellular diffusivity ($D_{\mathrm{in}}$) and signal fraction ($f_{\mathrm{in}}$). This conversion was done to simplify the presentation of the results but note that it cannot be done normally as $D_{\mathrm{in}}$ and $f_{\mathrm{in}}$ are unknown.

The performance of the representation was evaluated by investigating the effect of the total encoding time ($T_{\mathrm{max}}$) and maximum b-value ($b_{\mathrm{max}}$) on size and exchange estimates, knowing that increasing these tend to increase the level of violation of the assumptions underpinning the theory. To assess the performance of the representation, we studied how the bias and precision in parameter estimates as well as the effect size varied with $T_{\mathrm{max}}$ and $b_{\mathrm{max}}$. $T_{\mathrm{max}}$ was defined as the total duration of the longest-lasting waveform in a given protocol. Bias was defined as the difference between the parameter value expected from the simulations and the value obtained from the fitting. Precision was defined as the standard deviation across multiple fittings on data with uniquely generated noise. The effect size (Cohen's d) expected from a group study was calculated as

$$\mathcal{E} = \frac{\overline{X_1} - \overline{X_2}}{S_{X_1 X_2}}, \tag{23}$$

where $\overline{X_1} - \overline{X_2}$ is the difference in estimated means between two simulated groups and $S_{X_1 X_2}$ denotes the pooled standard deviation of the two groups.

To find the combination of $T_{\mathrm{max}}$ and $b_{\mathrm{max}}$ with the lowest bias, highest precision and largest effect size, we investigated protocols defined with $T_{\mathrm{max}}$ varying from 80 ms to 200 ms. At each $T_{\mathrm{max}}$, we applied the constraints $T \leq T_{\mathrm{max}}$, $\eta_k \leq 50\%$ and $\eta_d \leq 70\%$, which reduced the set of candidate waveforms to a compact region in the restriction-exchange space. We then chose four waveforms maximally separated on the convex hull of this region. The waveforms were scaled to achieve maximum b-values in the interval 0.5 to 10 ms/μm$^2$. At each combination of $T_{\mathrm{max}}$ and $b_{\mathrm{max}}$, signals were simulated for cylinder diameters of 2 and 5 μm and exchange rates of 0 and 5 s$^{-1}$. Effect size estimates were computed for these two settings using Eq. 23. Rice-distributed noise was added to the signals at SNR = $200 \cdot \exp(-T_{\mathrm{max}}/\mathrm{T}_2)$ where we assumed $T_{\mathrm{max}} = \mathrm{TE}$ and $T_2 = 80$ ms.

## 3.3 Investigating different microstructures for a fixed protocol

Our next aim was to analyse the performance of the representation under different microstructures but a fixed protocol. For each waveform family, a fixed test protocol was defined using waveforms selected from the restriction-exchange space after applying the constraints $T \leq T_{\mathrm{opt}}$, $\eta_{\mathrm{k}} \leq 50\%$ and $\eta_{\mathrm{d}} \leq 70\%$ where $T_{\mathrm{opt}}$ is the optimal maximum encoding time that gave the lowest bias, highest precision and largest effect size in the preceding section. Four waveforms were selected such that they were maximally separated on the convex hull of the region of waveforms satisfying the named constraints. Where possible, the waveforms were chosen such that one pair would strongly encode restriction (maximum separation in $V_\omega$) while the other one strongly encoded exchange (maximum separation in $\Gamma$). The waveforms in each protocol were scaled in amplitude to attain the maximum b-value $b_{\mathrm{opt}}$ found in the previous section.

The performance of the representation was evaluated by analysing the bias, crosstalk, and precision. Crosstalk was defined as the correlation between size and exchange rate estimates, for example, a size-dependent bias in the exchange rate. For this analysis, signals were simulated using the fixed protocols and all combinations of cylinder diameter and exchange rate from the sets $d = 2, 3, 4, \dots, 20$ μm and $k = 0, 1, 2, \dots, 20$ s$^{-1}$. All other simulation parameters were kept fixed at $f_{\text{in}} = 1 - f_{\text{ex}} = 0.7, D_{\text{in}} = 1.2$ μm$^2$/ms, $D_{\text{ex}} = 1.2$ μm$^2$/ms. Precision was assessed by adding Rice-distributed noise at a generous SNR of 200 (at b = 0) to signals simulated for cylinders with two diameters (4 and 8 μm) at an exchange rate of 2 s$^{-1}$ and two exchange rates (2 and 10 s$^{-1}$) at a diameter of 8 μm. The fitting procedure was in all cases the same as the one described in section 3.2.

### 3.4 Visualising common protocols in the restriction-exchange space

Finally, in order to understand how previously published protocols for probing of restricted diffusion and/or exchange manifest in our restriction-exchange space, we computed and visualized Γ and $V_\omega$ for the waveforms used in FEXI [39], IMPULSED [28], OGSE [66], SDE optimised for estimation of Kärger model parameters [67] and correlation tensor imaging[68] (CTI). Note that only the part of the CTI protocol featuring parallel gradient pairs is considered here. Table 1 summarizes the parameters of the five protocols.

# 4 Results

## 4.1 Characterization of waveforms in the restriction-exchange space

Our analysis of a large set of waveforms showed that these form a compact region in the restriction-exchange space, here visualized in terms of $V_\omega^{-1/2}$ and Γ (Fig. 1). In the SDE case (panel A), narrow pulses gave strongest restriction encoding (shortest $V_\omega^{-1/2}$). An increase in the diffusion time led to an elevated exchange encoding (longer Γ). Increasing both the pulse width and the diffusion time led to stronger exchange encoding but weaker restriction encoding. In the FWF case (panel B), different types of waveforms clustered into subregions populated by waveforms resembling SDE, DDE or OGSE. The left edge was exclusively populated by SDE-like waveforms. Such waveforms are the most efficient and high efficiency was needed in the left edge in order to deliver the required b-value at low Γ, for which the total encoding time is short. SDE-like waveforms also yielded the lowest $V_\omega$ due to their lack of oscillations. At longer Γ, DDE-like waveforms were more common on the bottom of the region (high $V_\omega$) as DDE oscillates more than SDE. The region peaked at a Γ-value of 50 ms due to the condition imposed on the maximum encoding time (250 ms). The right half of the triangular region comprised only oscillating waveforms, because oscillations result in an exchange-weighting function ($q_4$) with non-zero values at longer times which gives a longer Γ. Oscillations also add power at high frequencies of the encoding spectrum, which results in higher $V_\omega$. Taken together, these factors explain the increase in $V_\omega$ with increasing Γ to the right of the peak.

Figure 2 shows the restriction-exchange space for four types of waveforms: SDE, DDE, OGSE and FWF together with the outlines showing where waveforms comply with assumptions in terms of exhibiting low values of $\eta_d$ and $\eta_k$. Given any combination of constraints, FWF provides a wider range of candidate waveforms than SDE which in turn provides a wider range than both DDE and OGSE. The reason the regions made up by DDE and OGSE candidates are notably smaller than those made up by SDE or FWF candidates is that the former cannot achieve high b-values at short encoding times. Additionally, they can only provide strong exchange- and restriction-weighting due to their oscillatory behaviour. When considering the demands on $\eta_d$ and $\eta_k$, we can see that imposing stricter constraints on $\eta_d$ results in a total exclusion of DDE and OGSE candidates. This occurs because such waveforms have encoding power at higher frequencies (this is also evident in Fig. A3 from Appendix A). More stringent demands on $\eta_k$ exclude longer waveforms and constrains the set of candidates in a somewhat predictable manner by excluding the rightmost part of the waveform region.

## 4.2 Study of different protocols given fixed microstructure parameters

To understand how the set of waveforms that make up a protocol influences performance in terms of parameter estimation, we evaluated the effect of the maximum encoding time ($T_{\max}$) and maximum b-value ($b_{\max}$) on bias, precision, and effect size. Results are shown in Fig. 3 for waveforms from the SDE family. Largely similar results were obtained using FWF waveforms (see supplementary material, Fig.D1). Panel A shows the size-related results for protocols with variable $T_{\max}$. There is a general increase in bias, decrease in precision and decrease in effect size with increasing $T_{\max}$. The favoured range of values of $T_{\max}$ is about 90 – 120 ms. Panel B shows size-related results for variable $b_{\max}$. There is a decrease in bias and increase in precision with increasing $b_{\max}$ up to 5 ms/μm$^2$, above which the curves tend to flatten out. The effect size generally grows with $b_{\max}$. Regarding exchange estimates, panel C shows a decline in bias up to a $T_{\max}$ of about 120 ms where the bias becomes independent of $T_{\max}$. Both precision and effect size peak around this same $T_{\max}$ (120 ms). In Panel D, the minimum bias, maximum precision and largest effect size occur around a $b_{\max}$ of 5 ms/μm$^2$.

Two important concepts are illustrated by Figure 3. Increasing the maximum encoding time increases the range of $\Gamma$ that a given protocol can achieve, and due to increased leverage, the precision in estimates of exchange rate is improved. However, increasing the maximum encoding time also prolongs the echo time which in turn reduces SNR. Eventually, the decrease in SNR overpowers the benefit of increased leverage and overall performance declines. A similar argument can be made regarding the maximum b-value.

Figure 4 shows one FWF and one SDE protocol selected for further studies. Both protocols feature four different waveforms, a total encoding time of 120 ms, a maximal b-value of 5 ms/μm$^2$ and a maximum $\eta_k$ and $\eta_d$ of 50% and 70%, respectively (no waveforms fulfilled these criteria for DDE and OGSE). The FWF protocol comprised two waveforms with the same restriction-weighting but different exchange-weighting (waveform 1 and 3) and two waveforms with equal exchange-weighting but different restriction-weighting (2 and 4). The rationale behind this configuration is that the first pair of waveforms maximizes exchange-driven contrast while the second pair maximizes restriction-induced contrast. In the SDE case, this cross-like configuration could not be achieved, so the first waveform pair (1 and 3) features both variable exchange and restriction weighting.

Figure 5 elucidates the mechanisms explored in this study using the FWF protocol. Panel A shows the second cumulant evaluated (using Eq. 4) for cylinder diameters in the range 2 to 20 μm and panel B shows the encoding power spectrum for each waveform in the protocol alongside the diffusion spectrum for a cylinder with a diameter of 10 μm. Waveforms 1 and 3 have the same $V_\omega$ and – as predicted – yield the same cumulant for diameters below 10 μm. For larger diameters, the assumptions of the representation are no longer respected and the second cumulants diverge for waveforms 1 and 3. As the restriction encoding is no longer described by $V_\omega$ alone, changes in diameter may erroneously be interpreted as changes in exchange rate. Panel C shows the fourth cumulant (Eq. 16) for microstructures defined by a diameter of 10 µm and exchange rates varying from 0 to 20 $s^{-1}$. Panel D shows the exchange-weighting function ($q_4$) for the four waveforms. As expected from Fig. 4, waveforms 1 and 3 shows the maximal separation in the fourth cumulant, while waveforms 2 and 4 show an intermediate and overlapping value. Contrary to the case for restricted diffusion, waveforms selected to exhibit equal exchange encoding yield the same value regardless of the ground truth exchange rate.

## 4.3 Protocol performance for different microstructures

Figure 6 shows the evaluation of bias and crosstalk for the SDE and FWF protocols. There is a general increase in bias with increasing size (panels A and E). This is expected since larger structures challenge the assumptions of the representation. Concerning crosstalk between varying diameter and estimated exchange rate, Fig. 6 suggests that FWF outperforms SDE as it demonstrates no correlation between the two parameters up to a diameter of 7 µm whereas there is crosstalk at all levels for SDE (panels B and F). This result is likely a manifestation of the restriction-exchange configurations presented in Fig. 4, where the cross-like configuration available for FWF has a noticeable advantage. Panels C and G show the bias in exchange estimates for variable exchange rates evaluated for the FWF and SDE protocols. The bias increased with increasing exchange rate. This result stems from two sources. The first is a violation of the time-domain approximation made in the definition of the exchange-weighting time. The second is the assumption of barrier-limited exchange which becomes increasingly void with increasing exchange rate. The transition from non-barrier-limited to barrier-limited is depicted in Fig. E1 (see Supplementary material) where slow exchange is barrier-limited for a wider range of sizes. Fast exchange is associated with a violation of the barrier-limited assumption except for small sizes. Combinations of large sizes and fast exchange constitute a gross deviation from the assumptions of the theory, leading to the drastic decline in performance evident in panels C and G of Fig. 6. Crosstalk between size estimates and varying exchange rate was observed in particular for smaller diameters (panels D and H). The level of crosstalk (slope of the curves) was larger for SDE than the FWF protocol. Overall, Fig. 6 suggests that the proposed representation is applicable to barrier-limited exchange coupled with small sizes (2-7 µm). Note that the compartment size determines the maximum exchange rate than can be reliably estimated with our approach. For example, at a size of 7 µm, the theory is valid for exchange rates up to 10 $s^{-1}$. Smaller sizes imply that faster exchange rates can be accommodated (Fig. E1).

Figure 7 shows a comparison between the SDE and FWF protocols in terms of precision. The parameter distributions follow the same patterns for both protocols. The larger the value of the underlying parameter, the larger the bias in the estimates and the higher the precision. However, the precision in size estimates obtained with FWF protocol is about two times higher than that obtained with the SDE protocol (Fig. 7). This result can be attributed to the larger range of restriction-weightings ($V_\omega$) delivered by the FWF protocol (Fig. 4). This leverage is especially important at smaller sizes where restriction-induced signal contrast is lower. Regarding exchange, Fig. 7 shows that the FWF protocol also adds a noticeable improvement in precision compared to SDE. Taken together, the results of Fig. 7 suggest that using FWF can improve the precision of parameter estimates by up to a factor of 2.

### 4.4 Visualising common protocols in the restriction-exchange space

Figure 8 provides a visualisation of the restriction-exchange weighting properties of waveforms from literature, including FEXI, OGSE, IMPULSED, SDE optimised for exchange estimation and CTI. Here, FEXI demonstrates pronounced exchange sensitivity (variation in $\Gamma$) coupled with negligible restriction sensitivity (near constant $V_\omega$). OGSE is mostly sensitive to restriction, with a mild exchange sensitivity that tends to decrease with increasing frequency. IMPULSED strongly encodes restriction and is insensitive to exchange. The optimised SDE protocol is strongly sensitive to exchange but exhibits a greater restriction sensitivity than FEXI. CTI is sensitive to exchange (variable $\Gamma$) but not to restricted diffusion (constant $V_\omega$).

# 5 Discussion and conclusion

In this work, we proposed and evaluated an approach for analysis of restricted diffusion and water exchange in diffusion MRI. The approach features two aspects. The first is a signal representation that allows estimation of parameters connected to restricted diffusion and exchange. The second is the concept of mapping a gradient waveform to two scalar parameters that span the restriction-exchange space with one dimension for restricted diffusion [32] and another for exchange [42]. The utility of these aspects will be discussed below, starting with the second one.

Mapping waveforms into the restriction-exchange space provides useful insights. Here, we elaborate on three examples. First, it enables the selection of waveforms that maximise signal contrast due to either exchange or restriction, or both. For example, waveforms that lie on a horizontal line in the space (constant $V_\omega$ but varying $\Gamma$) provide variable sensitivity to exchange with constant sensitivity to restriction. Such waveforms would, in principle, enable specific estimation of exchange without bias from restricted diffusion – at least as long as the assumptions hold. Second, it shows that the set of available free waveforms span a triangular region in the restriction-exchange space, with different waveform types clustered (Fig. 1). Third, transforming waveforms to this space can be used to analyse and compare previous approaches. In this work, we considered five examples and found that the methods were variably specific to their intended purpose. For example, FEXI and an optimised SDE protocol [67] both lie on the $\Gamma$-axis of the restriction-exchange space, giving sensitivity to exchange but not restriction. Both OGSE and IMPULSED predominantly encode restriction. Low frequency OGSE exhibits some degree of exchange sensitivity which may confound the interpretation of parameter estimates. For instance, OGSE has been applied extensively for mapping restricted diffusion under the assumption of negligible exchange encoding [69–72]. The fifth and last example considered is CTI, which was recently proposed to disentangle restriction-driven intra-compartmental kurtosis (μK) from other kurtosis sources [68]. Interestingly, our analysis suggests that CTI (in one direction) has sensitivity to exchange, which may confound the estimation and interpretation of μK. Alternatively, one could argue that exchange is to be regarded as one source of μK.

The numerical evaluation highlighted intervals of validity of the proposed signal representation. With respect to exchange, estimates showed a uniform bias for smaller sizes where the barrier-limited assumption[40] is respected (Fig. E1 supplementary material). Small sizes in this context refers to sizes up about 7 µm. In this range, approach is applicable to systems with exchange rates up to 20 $s^{-1}$. This validity interval covers a subset of the conditions observed *in vivo.* For instance, previous work has reported exchange rates in the human brain in the range $1-40$ $s^{-1}$[73–76,40,77]. Furthermore, it should be noted that the effect of exchange on the diffusion-weighted signal can manifest from other sources than membrane permeability. An apt example is structural disorder which has been hypothesised to compete with water exchange in human grey matter [78].

In the case of restricted diffusion, the interval of validity covered sizes between approximately 2 and 7 µm. For smaller sizes, the underlying compartment diameter approaches the resolution limit and is therefore inherently challenging to map without bias [32,79]. Below this limit, our size estimates also become dependent on the exchange rate. Accurate mapping of axonal sizes is thus challenging with this approach, as these have diameter distributions with medians around 1 µm [80,81]. However, other approaches using ultra-strong gradients and custom experiments show promising results [82,83]. For sizes above the validity region, the restriction-weighting properties of a gradient waveform are no longer explained by $V_\omega$ alone (Fig. 5). This scenario corresponds to a mixture of the motional narrowing and free diffusion regimes discussed in previous work [32,84]. Note that this implies that waveforms that lie along the exchange axis of the restriction-exchange space (such as FEXI) may become sensitive to restricted diffusion in environments with large structures. Indeed, previous work has shown that restricted diffusion can bias exchange rate estimates [85], which we here refer to as crosstalk (Fig. 6). For diameters above the 7 µm threshold, microstructure models such as IMPULSED have been shown to yield more accurate estimates [28]. However, while such microstructure models may provide better size estimates than the representation presented here, their accuracy hinges on the validity of the model. In any event, any approach seeking to map only restricted diffusion must employ waveforms that provide equal sensitivity to exchange. A welcome future development would be a description of restriction and exchange encoded with arbitrary gradient waveforms, without relying on the assumptions of the present framework or assumptions on the microstructure.

The use of either a signal representation or a microstructure model, with terms used as defined in Novikov et al. [86], has implications on protocol design that warrant further discussion. The assumptions behind our signal representation led to a constraint in the restriction-exchange space which limited the set of waveforms available for designing a protocol (Fig. 2). The performance analysis also showed a surprisingly high optimal b-value of 5 ms/$\mu m^2$ coupled with a relatively short maximum encoding time of 120 ms which excluded oscillating gradient waveforms as a feasible design. A direct consequence is the inability to leverage the strong restriction encoding delivered by such waveforms. This trade-off between bias and precision is commonplace in other cumulant-expansion based frameworks such as diffusion tensor imaging and diffusion kurtosis imaging [87–89]. It is worth noting that, under all constraints placed on the restriction-exchange space, free waveforms exhibited greater versatility than other waveform types considered in this work—an expected result due to their flexibility of shape [13,51,64]. The benefit of using these waveforms was also highlighted by a two-fold improvement in precision compared to SDE. We expect that this advantage of FWF would only be amplified by a microstructure model compatible with all waveforms, but the development of such a model remains the target of future work.

Note that effects of anisotropy were not covered by the present work. This shortcoming is important to tackle because diffusion shows considerable directional dependence in biological tissue. The one-dimensional analysis we have presented here neglects effects of microscopic diffusion anisotropy—a limitation that has been addressed using tensor-valued diffusion encoding [90–92]. Another remedy is the multidimensional diffusion encoding framework developed by Lundell et al. [93] to disambiguate the effects of microscopic anisotropy and time-dependent diffusion. Previous work has shown that the exchange-weighting parameter $\Gamma$ is influenced by the gradient orientation in non-linear tensor encoding acquisitions [51]. A more complete forward model would account for anisotropic media and such a framework already exists for non-exchanging restrictions with the potential for extension to include exchange [60]. Incorporating these findings and refining the theory accordingly is reserved for future work.

In conclusion, our signal representation offers an intuitive and explicit description of how the phenomena of restricted diffusion and water exchange manifest in the signal. More importantly, it provides insights into the restriction- and exchange-weighting properties of gradient waveforms. The approach provides an important guide in designing an experiment to maximise and disentangle the two independent contrasts. Future work will develop a more complete forward model to address the limitations of the current theory. Particularly, future work will eliminate the approximations made in this work and transform the theory from one dimension to a tensorial description.

# Figures

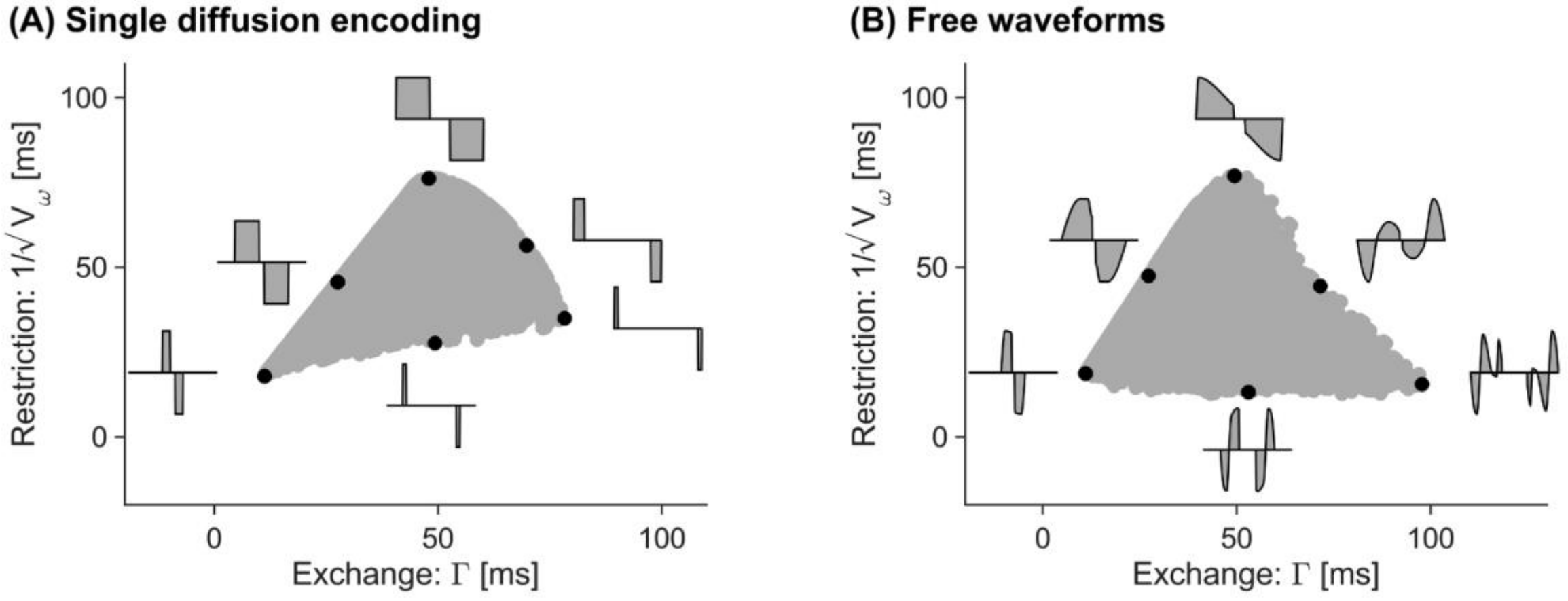

Figure 1: The restriction-exchange space for all available waveforms from two families: SDE (A) and FWF (B). The space is defined by Γ and $V_\omega$, which quantify the strength of exchange and restriction weighting for a given gradient waveform. Note that the y-axis in this figure shows the inverse square root of $V_\omega$ to obtain the same unit and scale on both axes. Black points show the position of waveform examples, which were all plotted on the same time axis with a maximal duration of 250 ms. Waveforms were for both SDE and FWF generated with a maximum encoding time of 250 ms, a maximum gradient amplitude and slew rate of 80 mT/m and 70 T/m/s generating a minimum b-value of 5 ms/μm$^2$. Each point represents the most efficient waveform out of a series of candidates with similar restriction- and exchange-weighting. Note that the analysis shown here uses effective waveforms with an implied refocusing pulse in the middle.

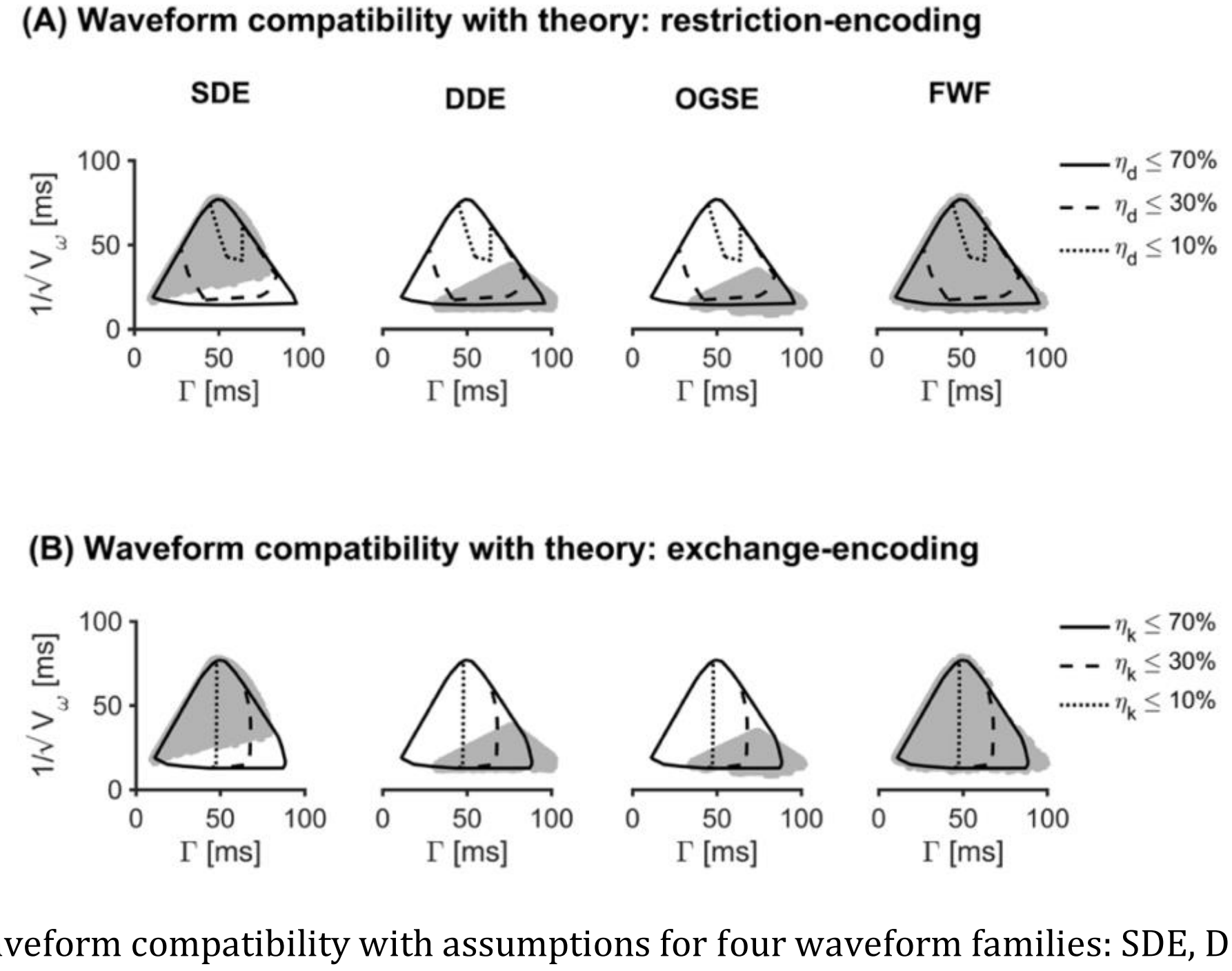

Figure 2: Waveform compatibility with assumptions for four waveform families: SDE, DDE, OGSE and FWF. Panel A shows how well waveforms comply with the restriction-related assumption as quantified by $\eta_d$ (Eq. 21). As the requirement is sharpened (lower $\eta_d$), the number of waveforms complying with it gradually shrinks (solid, dashed and dotted lines). Panel B shows the corresponding analysis for the parameter $\eta_k$ (Eq. 22), which quantifies how well waveforms comply with the assumption behind the theory. A sharper waveform requirement (smaller percentage) is associated with a smaller region.

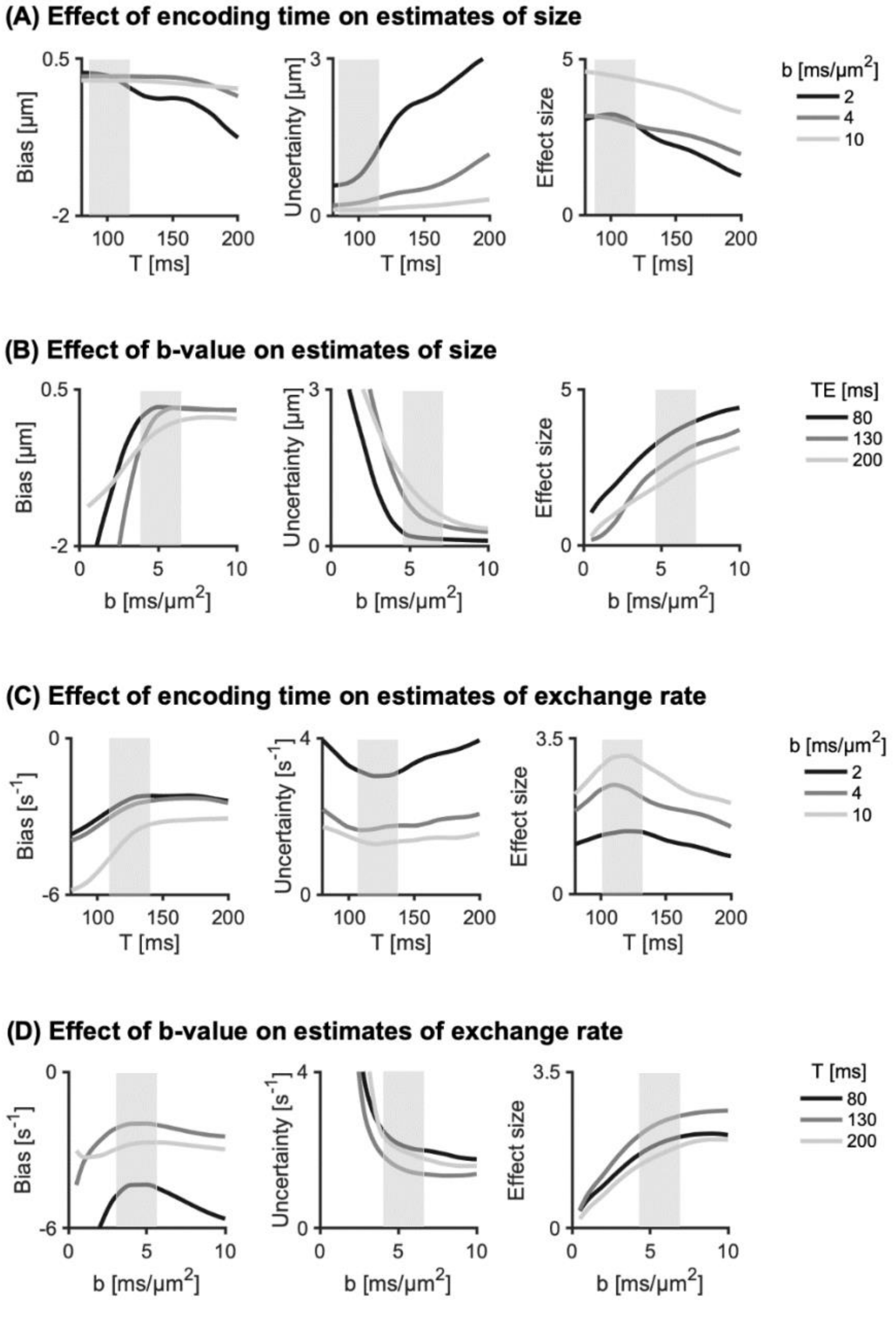

Figure 3: Bias, precision and effect size in compartment size and exchange rate as a function of maximum encoding time ($T_{\max}$) and maximum b-value ($b_{\max}$). Precision is quantified by the standard deviation (labelled "uncertainty" in the figure). SNR was set to 200 at TE = 0 and b = 0 and a $T_2$ relaxation time of 80 ms was assumed. The grey bars on each subplot mark the desired range of values of $T_{\max}$ or $b_{\max}$. In Panel A, the favoured range of values of $T_{\max}$ (least bias, highest precision and largest effect size) is 90-120 ms. Panel B shows that the favoured $b_{\max}$ lies around 5 ms/μm$^2$. Applying the same procedure to the exchange-related results in panels C and D, the desired values of $T_{\max}$ and $b_{\max}$ are around 120 ms and 5 ms/μm$^2$. Collectively, this figure suggests an optimum where $T_{\max}$ is approximately 120 ms and $b_{\max}$ approximately 5 ms/μm$^2$.

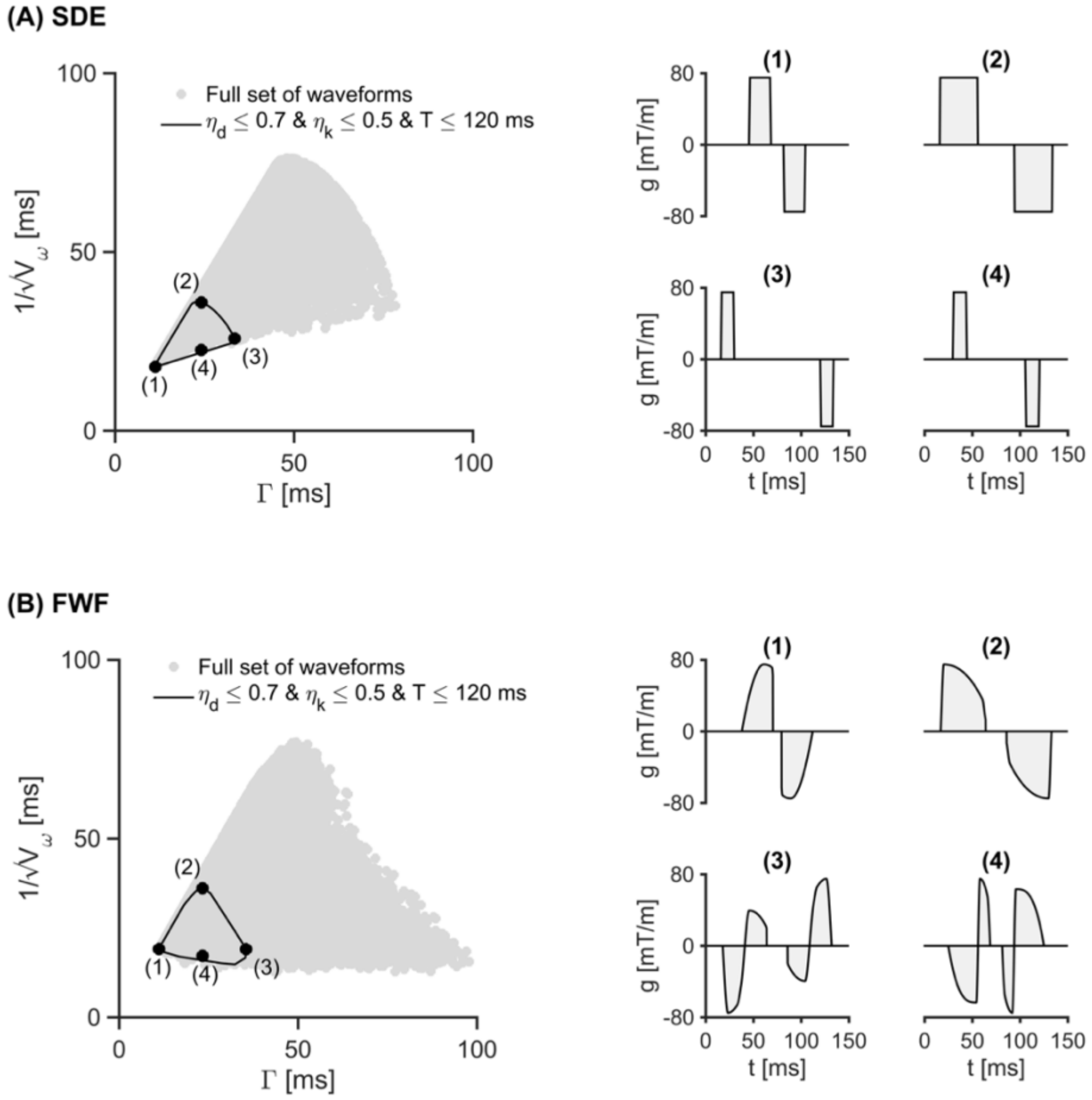

Figure 4: The final SDE and FWF protocols comprised four waveforms. These were selected among waveforms with a maximum encoding time of 120 ms, gradient amplitude and slew rate of 80 mT/m and 70 T/m/s, respectively, which were found within the region marked by a black line. The black dots mark the positions of the selected waveforms. For the SDE protocol, the $(\Gamma, V_\omega^{-1/2})$ values for waveforms 1 to 4 are, in order, (11, 17), (24, 35), (33, 25) and (24, 22) ms. For FWF, these numbers are (11, 19), (23, 36), (35, 19) and (23, 17) ms. Thus, waveforms 1 and 3 have the same values of $V_\omega$ but different Γ and waveforms 2 and 4 have the same Γ but different $V_\omega$. The range of Γ values between waveforms 1 and 3 is 24 ms for FWF and 22 ms for SDE. Forcing waveforms 1 and 3 to have the same $V_\omega$ for SDE reduces this range to 17 ms.

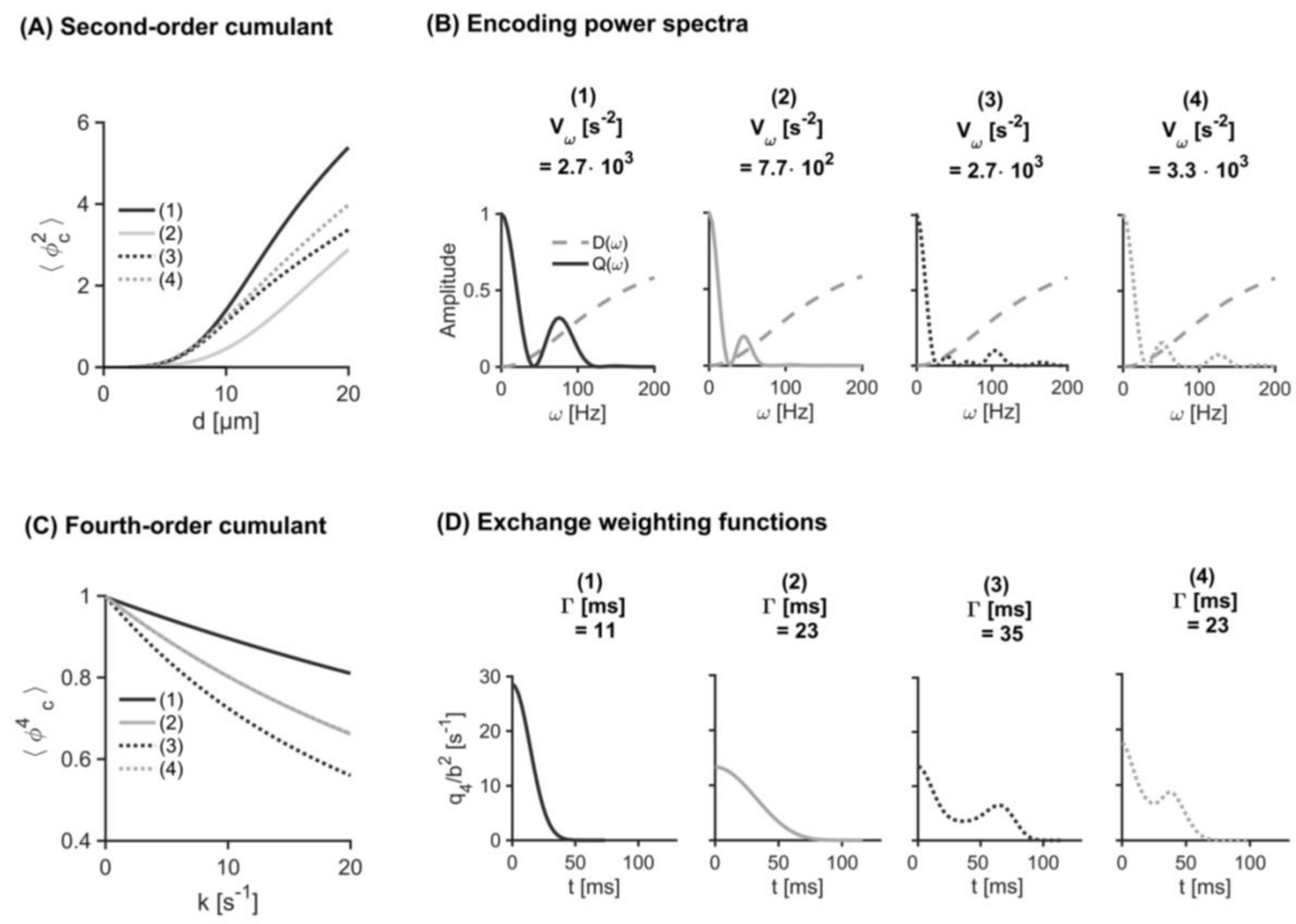

Figure 5: Illustration of the restriction- and exchange encoding performed by the free waveforms in Fig. 4. In panel A, waveforms 2 and 4 have the weakest and strongest restriction weighting, respectively. This follows from that the power spectrum of waveform 4 contains power at higher frequencies than that of waveform 2. Waveforms 2 and 4 have equal exchange weightings and should therefore show no exchange-driven contrast. Indeed, their fourth-order cumulants in panel D coincide. In panel B, waveforms 1 and 3 have the weakest and strongest exchange-weighting, respectively. This is evident from the $q_4$ functions in panel B, where waveform 3 features non-zero values at longer times than waveform 1. Their restriction weightings are equal, which means that they show no contrast with varying diameter. Panel C shows that waveforms 1 and 3 begin to exhibit pronounced contrast when the diameter grows above 10 μm.

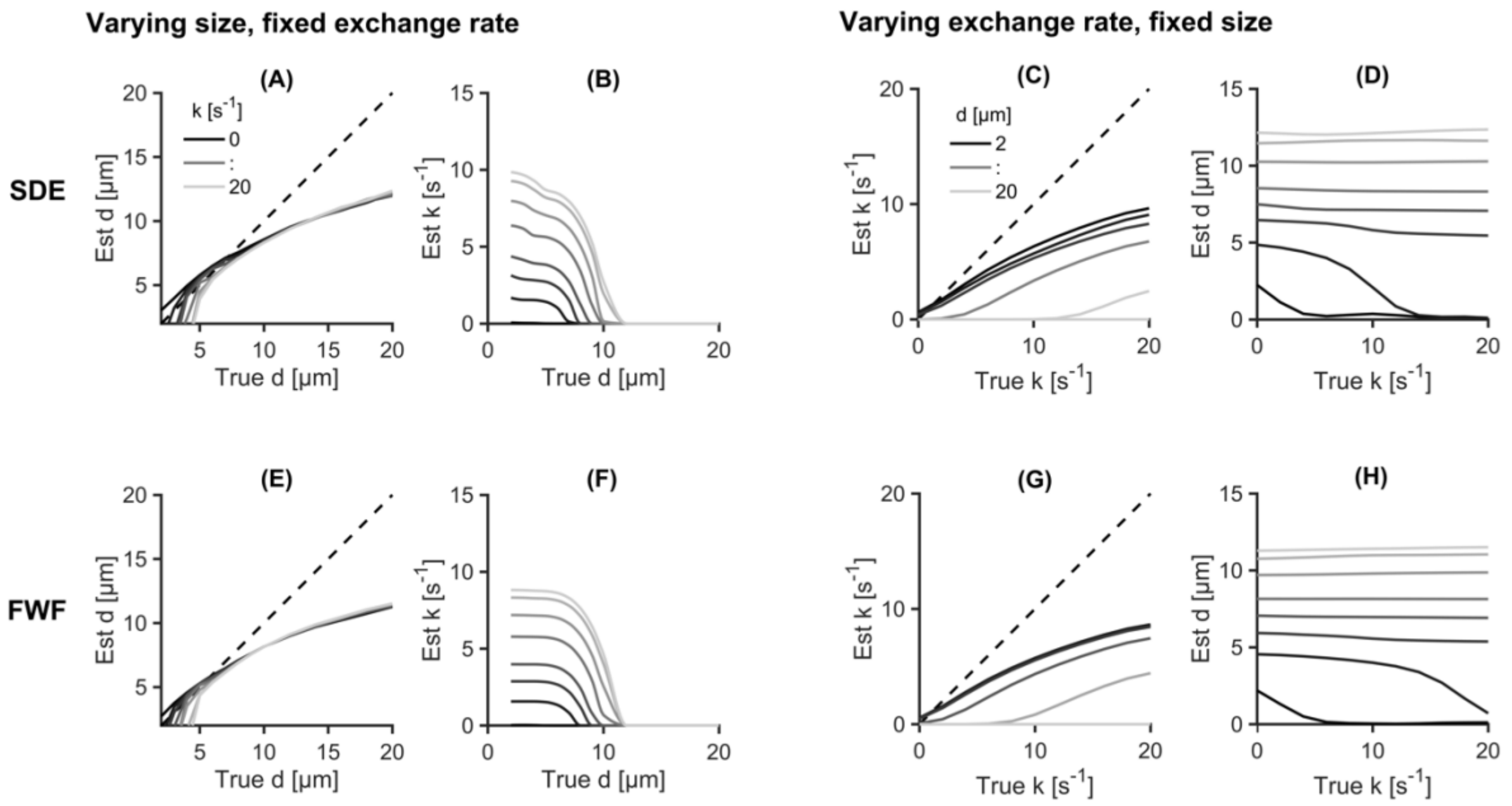

Figure 6: Bias and crosstalk in estimates of size and exchange rate. Plots (A) and (E) show the bias in size estimates, which increases with size. Varying size has no effect on exchange estimates up to a size of 7 μm, after which there is strong correlation between the two quantities (B and F). FWF shows an advantage over SDE in this regard. Plots (C) and (G) show that the bias in exchange rate increases with exchange. Plots (D) and (H) present the crosstalk between size and exchange. Both FWF and SDE show similar trends, with the size faithfully captured at all exchange rates except for the smallest sizes. Overall, Fig. 6 shows that the proposed representation is applicable for barrier-limited exchange and small sizes (2-7 μm).

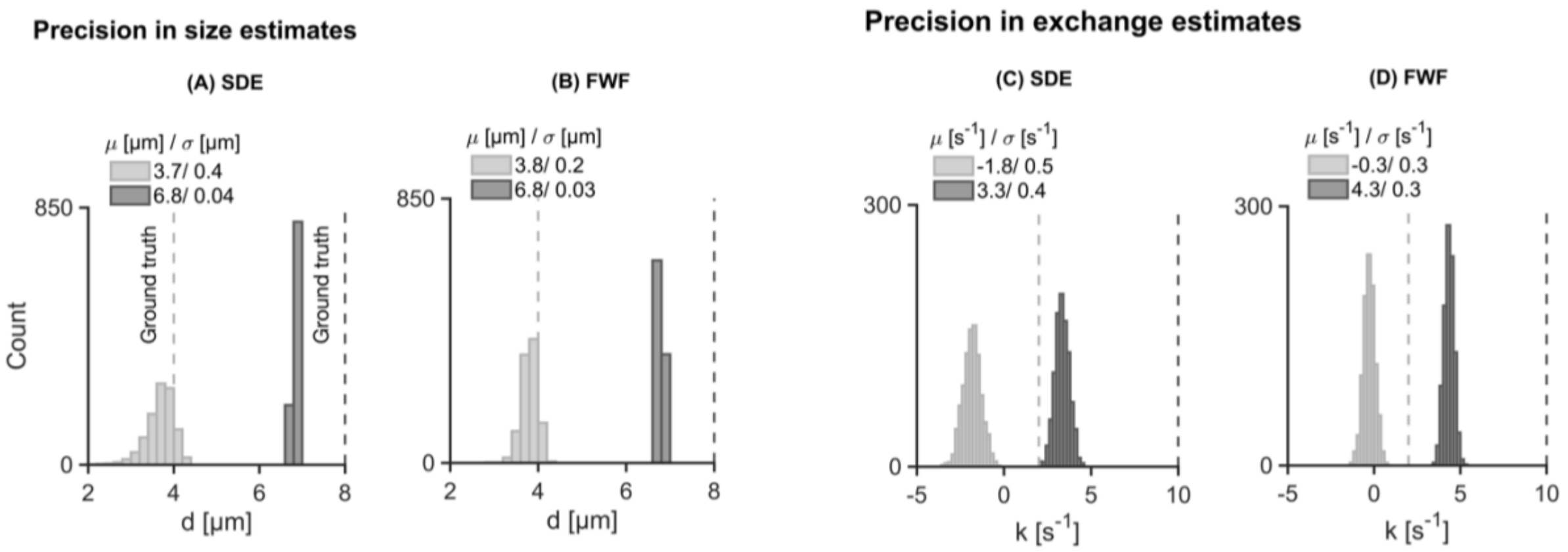

Figure 7: Precision in estimates of compartment size and exchange rates—a comparison between single diffusion encoding (SDE) and free waveforms (FWF). The ground truth sizes are 4 and 8 μm at a fixed exchange rate of 2 $s^{-1}$. For exchange, the ground truth values are 2 and 10 $s^{-1}$ at a fixed size of 8 μm. Symbols $\mu$ and $\sigma$ represent the mean and standard deviation of estimates obtained at an SNR of 200 (at b = 0). Large diameters/exchange rates result in larger bias but higher precision than smaller diameters/exchange rates. For size estimates, FWF gives twice the precision of SDE. There is also some improvement in precision in exchange estimates with FWF compared to SDE.

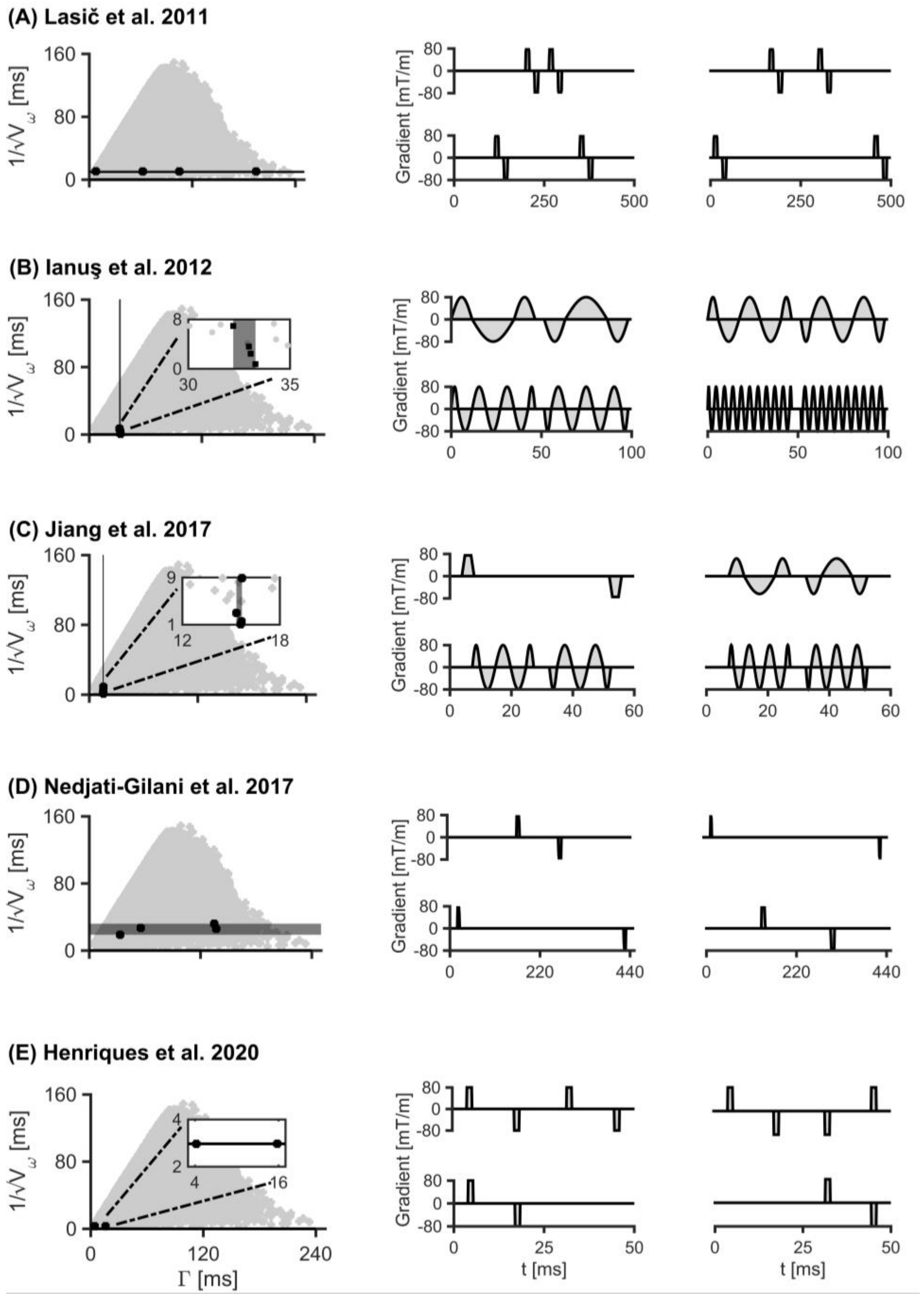

Figure 8: Visualisation of five different protocols on the restriction-exchange space: FEXI, OGSE, IMPULSED, SDE optimised for exchange estimation and correlation tensor imaging (CTI). Note that only four out of eight waveforms from the CTI protocol are shown because the rest are a rescaling of those shown here, which does not change Γ and $V_{\omega}$. The restriction-exchange space (grey triangle on the left) was generated using free waveforms with a maximum gradient strength of 80 mT/m, slew rate of 70 T/m/s and a maximum total encoding time of 500 ms. Each of the five protocols considered here comprise four waveforms shown adjacent to the restriction-exchange space. FEXI lies along the exchange axis of the space, which means that the protocol is sensitive to exchange but not restriction. OGSE is sensitive to restriction but exhibits minor exchange sensitivity at low frequencies (<50 Hz). IMPULSED is largely sensitive to restriction and shows negligible exchange sensitivity. The optimised SDE protocol is strongly sensitive to exchange but demonstrates greater restriction sensitivity than is seen in FEXI. CTI (assuming linear encoding) predominantly encodes exchange.

# Tables

Table 1: Gradient waveform parameters for five different protocols.

| Protocol | Parameters |
|---|---|
| FEXI [39] | $\delta_f = 11$ ms; $\delta_d = 9$ ms; $t_f = 21$ ms; $t_d = 22$ ms;<br>$t_m = [30, \quad 100, \quad 200, \quad 410]$ ms |
| OGSE [66] | $f = [20, \quad 40, \quad 60, \quad 200]$ Hz<br>pause duration: $(\Delta - \delta) = 5$ ms; $\delta = 50$ ms |
| IMPULSED [28] | $\delta_{SDE} = 4$ ms; $\Delta_{SDE} = 48$ ms<br>$\delta_{OGSE} = 20$ ms; $\Delta_{OGSE} = 25$ ms; $f_{OGSE} = [50, 100, 150]$ Hz |
| SDE [67] | $\delta = [7, \quad 3.9, \quad 5.7, \quad 9.4]$ ms;<br>$\Delta = [102, \quad 412, \quad 406, \quad 169]$ ms |
| CTI [68] | $\delta = 1.5$ ms; $\Delta = 13$ ms; $\mathrm{t_m} = 13$ ms<br>$(\lvert q_1 \rvert, \lvert q_2 \rvert) = [(q_b, q_b); (q_b, q_b); (q_b, 0); (0, q_b)]$<br>$q_b = \sqrt{\left(b_{max}/(\Delta - \delta/3)\right)}$ |

$\delta_f/\delta_d$: filter/detection block pulse width; $t_f/t_d$: filter/detection block diffusion time; $t_m$: mixing time